\documentclass[twoside]{ilcws07}
\usepackage[latin1]{inputenc}
\usepackage[dvips]{graphicx,epsfig,color}
\usepackage{wrapfig,rotating}
\usepackage{amssymb,amsmath,array}

\begin{document}
\title{
A Study of $e^+e^- \to H^0A^0$ Production at 1~TeV \\ 
and the Constrain on Dark Matter Density} 
\author{Marco Battaglia$^{1,2}$,
Benjamin Hooberman$^1$,
Nicole Kelley$^1$
\vspace{.3cm}\\
1- University of California - Dept of Physics, Berkeley, CA - USA \\
\vspace*{-.2cm}\\
2- Lawrence Berkeley National Laboratory, Berkeley, CA - USA \\
}

\maketitle

\begin{abstract}

\end{abstract}

\section{Introduction}

The connections between Cosmology and Particle Physics through Dark Matter 
(DM) have received special attention in the last few years for sharpening 
the physics case of collider physics at the TeV frontier. 
There are many extensions of the Standard Model (SM), which include a new, 
stable, weakly-interacting massive particle, possibly responsible for the observed 
relic DM in the Universe. The LHC will provide first important data to address the 
question whether one of these scenarios is indeed realised in nature. The ILC 
measurements of the properties of a DM candidate and of those other particles 
participating to its interactions in the early Universe may allow us to predict 
its relic density with an accuracy comparable to that currently achieved by CMB 
observations at satellites. With these data in hand, the comparison of the results 
would have striking consequences for our understanding of dark matter.

\section{Neutralino Dark Matter Density in MSSM and the ILC}

Supersymmetry emerges as the best motivated theory of new physics beyond
the SM. It solves a number of problems, intrinsic to the SM and, most important to our 
discussion, the conservation of R-parity introduces a new stable, weakly interacting 
particle. The WMAP CMB data, and other astrophysical data, already set rather stringent 
bounds on Supersymmetry parameters, if the neutralino is responsible for
saturating the amount of DM observed.

The potential of ILC data at 0.5~TeV and 1.0~TeV for determining the DM relic 
density, $\Omega_{\chi}$, in Supersymmetry has been investigated in detail 
in~\cite{Baltz:2006fm}. This study selected a set of benchmark points, the so-called 
LCC points, representative of various scenarios and determined the $\Omega_{\chi}$ 
probability density function by a scan of the full MSSM parameter space and retaining 
those points compatible with the measurements available at the LHC and ILC within their 
accuracy.
 
\section{$e^+e^- \to H^0A^0$ at LCC-4 with Full Simulation}

We consider here a specific Supersymmetric scenarios, in which the DM candidate 
is the lightest neutralino, $\chi^0_1$ and its relic density is controlled by the 
rate of neutralino annihilation through the CP-even heavy Higgs pole $\chi \chi \to A$.
The LCC-4 benchmark point~\cite{Baltz:2006fm} is defined in the cMSSM, 
corresponding to the parameters $m_0$=380.00~GeV, $m_{1/2}$=420~GeV, 
$\tan \beta$=53, $A$=0, $Sgn(\mu)$=+1 and $M_{top}$=178~GeV. We use 
{\tt Isasugra 7.69}~\cite{Paige:2003mg} to compute the particle spectrum and we get
$M_{A^0}$=419.4~GeV, $M_{\chi^0_1}$=169.1~GeV and $M_{\tilde{\tau_1}}$=195.5~GeV. 
The $e^+e^- \to H^0A^0 \to b \bar b b \bar b$ process at $\sqrt{s}$ = 1~TeV ILC 
has already been studied for LCC4~\cite{Battaglia:2004gk}.
That study, based on the parametric detector simulation program {\tt Simdet 4.0},
showed that the $A^0$ boson mass can be determined to $\pm$0.8~GeV by imposing the 
natural width $\Gamma_A$ or to $\pm$2.0~GeV by a simultaneous fit to mass and width. 
These results, when combined with other measurements to be peformed at 0.5~TeV, 
allow us to predict the neutralino contributoion to the dark matter density in the 
Universe, $\Omega_{\chi}$ to a relative accuracy of 18~\% in generic MSSM scenarios.
  
Here, we repeat the same study on {\tt Geant-4}-based simulation~\cite{Agostinelli:2002hh} 
of the detector response and reconstruct the physics objects using processors developed
in the {\tt Marlin} framework~\cite{Gaede:2006pj}. This study adopts the LDC detector 
concept, which employs a large continuous gaseous tracker Time Projection Chamber 
surrounded  by a highly granular calorimeter and complemented by a high resolution 
Vertex Tracker, for which we have chosen the option based on CMOS monolithic pixel sensors.
The LDC detector is discussed in details elsewhere\cite{ldc}, the design is 
optimised for achieving excellent parton energy measurements, through the particle
flow algorithm, and precise extrapolation of particle tracks to their production 
point. Both these features are important to the analysis, which aims to suppress 
backgrounds by exploiting its signature 4-$b$ jet final state and requires good 
determinaton of energy and direction of hadronic jets to maximise the resolution on 
di-jet invariant masses. 
The jet energy resolution has been studied using a simulated 
sample of single $b$ jets in the energy range from 10~GeV to 210~GeV over a polar 
angle, $\theta$,  range $0.4<\theta<\pi/2 $, we get 
$\delta E/E = {\mathrm{(0.34\pm0.02)}}/\sqrt{E} \oplus {\mathrm{(0.015\pm0.005)}}$, 
which is consistent with the particle flow performance specifications.
Jet flavour tagging is based on three observables: the probability for all the particle 
tracks to originate at the event primary vertex, the fraction of the jet energy carried 
by secondary particles and the  $p_t$-corrected mass of the secondary particles.
These are combined to form a discriminant variable which peaks at one for $b$ jets 
and peaks at zero for non-$b$ jets.
\begin{wrapfigure}{r}{0.6\columnwidth}
\centerline{\includegraphics[width=0.59\columnwidth]{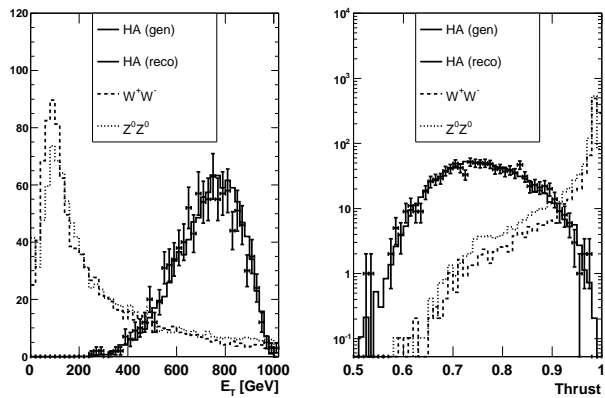}}
\caption{Tranverse energy and thrust distributions for $HA$, $Z^0Z^0$ and $W^+W^-$. 
Generator level distributions are plotted as histograms, results of {\tt Mokka + Marlin}
simulation and reconstruction are given for the signal process as points with error bars.}
\label{fig:cuts}
\end{wrapfigure}
At the chosen working point, an efficiency for 
$b$ jets of 85~\% is obtained with sufficient rejection of lighter quarks to effectively 
suppress the remaining non-$b$ background.
Signal events have been generated with {\tt Pythia 6.205}+{\tt Isajet 7.69}, 
including bremsstrahlung effects. These events have been passed through the 
full LDC simulation using the {\tt Mokka 06-03} program~\cite{Musat:2004sp} based 
on {\tt Geant-4}. The {\tt lcio}~\cite{Gaede:2003ip}
collections produced by {\tt Mokka} have been used as input for the {\tt Marlin} 
reconstruction. 
At $\sqrt{s}$ = 1~TeV, the effective $e^+e^- \to H^0 A^0$ production cross section, 
accounting for beamstrahlung, is 1.4~fb and the decay BR($A^0 \to b \bar b$) is 0.87. 
The main backgrounds are $Z^0 Z^0$, $W^+ W^-$ production and the inclusive $b \bar b b \bar b$ 
production. Their cross sections are 0.2~pb, 3.2~fb and 5.1~fb respectively.
We assume to operate the ILC at 1~TeV for a total integrated luminosity of 2~ab$^{-1}$.
Backgrounds can be significantly suppressed using event shape and kinematic variables.
We require events to fulfill the 
following criteria: total recorded energy in the event $E_{TOT}>$ 850~GeV, 
total transverse energy $E_{T}>$350~GeV, charged energy in the event $E_{CHA}>$ 350~GeV, 
number or reconstructed particles $N_{TOT}>$50, number of charged particles $N_{CHA}>$25, 
event thrust $<$.95 and $Y_{34}<$0.0025, where $Y_{34}$ is the 3 to 4 jet crossover value 
of the jet clustering variable. The distributions of some of these variables 
is shown in Figure~\ref{fig:cuts}) for backgrounds and signal, for which a comparison of 
the generator-level and reconstructed values is also given. 

After event selection, the di-jet pairing which minimises the di-jet mass difference has 
been chosen. The di-jet mass resolution has been improved by applying a 4-C fit.
\begin{figure}[hb!]
\centerline{\includegraphics[width=0.75\columnwidth]{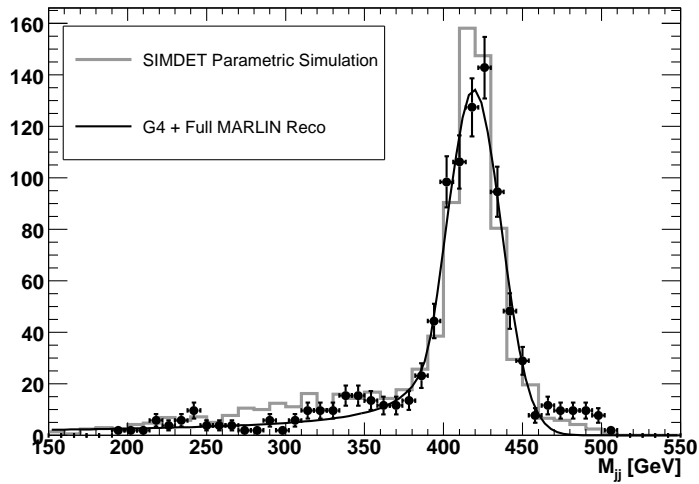}}
\caption{Dijet invariant mass distribution for $e^+e^- \to H^0A^0$ events selected by the 
analysis cut. Mass constraint fit and jet flavour tagging has been applied. The distribution 
for fully simulated and reconstructed events (points with error bars) is compared to that 
obtained with parametric simulation (histogram).}
\label{fig:jjmass}
\end{figure}
We have ported the {\tt PUFITC} algorithm, developed for the DELPHI experiment at LEP, 
into a dedicated {\tt Marlin} processor. The algorithm adjusts the momenta of the jets 
given by $\vec{p}_{F} =e^{a}\vec{p}_{M}+b\vec{p}_{B}+c\vec{p}_{C}$ where $\vec{p}_{F}$ 
is the fitted momentum, $\vec{p}_{M}$ is the measured momentum,$\vec{p}_{B}$ and 
$\vec{p}_{C}$ are unit vectors transverse to $\vec{p}_{M}$ and to each other, and 
$a$, $b$ and $c$ are the free parameters in the fit.  
The adjusted momenta satisfy 
a set of constraints while minimizing the fit $\chi^2$, which is given by 
$\Sigma_i$ $(a_i-a_0)^2/\sigma_a^2 + b_i^2/\sigma_b^2 + c_i^2/\sigma_c^2$, where $a_0$ 
is the expected energy loss parameter, $\sigma_a$ is the energy spread parameter and 
$\sigma_b$,$\sigma_c$ are the transverse momentum spread parameters. In this analysis 
we use the following constraints: $p_x=p_y=0$ and $E\pm|p_z|=\sqrt{s}$, where the last
condition accounts for beamstrahlung along the beam axis, $z$.
We report here preliminary results from the analysis of a sample of 1050 fully simulated 
signal events.
After applying final selection and mass constrained fit, the sample of events 
in the region 150~GeV$< M_{jj} <$550~GeV gives a selection efficiency of 23~\% for
signal $b \bar b \bar b$ decays. 
The resulting mass distribution is shown in Figure~\ref{fig:jjmass}. 
We describe the signal as a CRYSTAL BALL (CB) function~\cite{Skwarnicki:1986xj} and 
extract the $A^0$ mass, $M_A$, and width, $\Gamma_A$ have been by a multi-parameter fit 
leaving the CB parameters free. We determine the $A^0$ mass as (419.1$\pm$0.9)~GeV.
This result is remarkably close to that obtained in the earlier analysis, based on parametric 
detector simulation. The production and analysis of fully simulated and reconstructed 
background samples is currently under way.

\section{Further Constraints on $\Omega_{\chi}$}

The constraints on LCC4 derived from mass measurements at the LHC and ILC, provide a 
prediction of the DM density in the Universe to a relative accuracy of 18~\% with a 
generic MSSM model. 
\begin{wrapfigure}{r}{0.60\columnwidth}
\centerline{\includegraphics[width=0.59\columnwidth]{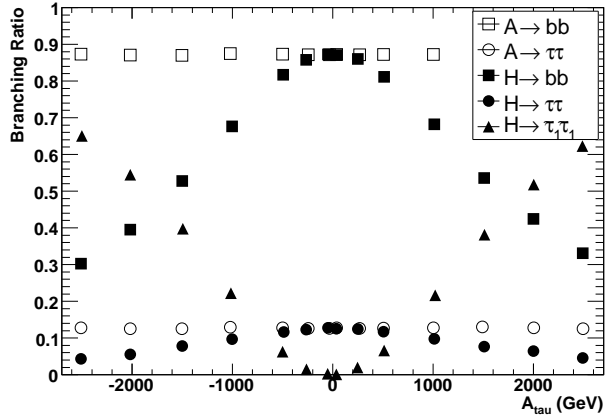}}
\caption{$H^0$ and $A^0$ decay branching fractions as a function of the 
stau trilinear coupling $A_{tau}$ predicted by {\tt HDECAY}. All the other 
MSSM parameters have been kept fixed to those corresponding to the LCC4 point.}
\label{fig:hdecay}
\end{wrapfigure}
This accuracy is still far from that achieved by CMB study with 
satellites. The main contribution to the remaining uncertainty is the weak constrain 
which the data provide to MSSM solutions where $\Omega_{\chi}$ is significantly lower 
than its reference for LCC4. A detailed study shows that these solutions are all
characterised by large values of the stau trilinear coupling, $A_{tau}$.
In the MSSM the $\tilde \tau$ coupling to the $H^0$ and $A^0$ bosons scales as 
$A_{tau} \frac{\cos \alpha}{\cos beta} + \mu \frac{\sin \alpha}{\cos \beta}$ and
$A_{tau} \tan \beta + \mu$, respectively.
In the funnel region the main annihilation mechanism is 
$\tilde \chi^0 \tilde \chi^0 \to A^0 \to b \bar b$ and $M_A < M_{\tilde \tau_1} + 
M_{\tilde \tau_2}$. The only $A^0$ decay into $\tilde \tau$s, allowed by CP symmetry, 
$A^0 \to \tilde \tau_1 \tilde \tau_2$ is kinematically forbidden.
At large values of $|A_{tau}|$, the stau decay process through the 
$H^0 \to \tilde \tau_1 \tilde \tau_1$ gets a sizeable branching fraction. This channel
contributes to the neutralino annihilation rate through $\tilde \chi^0 \tilde \chi^0 
\to H^0 \to \tilde \tau_1 \tilde \tau_1$, bringing down the corresponding relic density, 
as observed in the MSSM scans. At the same time, a determnation of the branching fraction 
of the decay $H^0 \to \tilde \tau_1 \tilde \tau_1$, allows to constrain $|A_{tau}|$. 
Figure~\ref{fig:hdecay} shows the decay branching fractions of the $A^0$ and $H^0$ bosons
computed using the {\tt HDECAY 2.0} program~\cite{Djouadi:1997yw} as a function of $A_{tau}$.
Now, a large $H^0 \to \tilde \tau_1 \tilde \tau_1 \to \tau \tilde \chi^0 \tau \tilde \chi^0$ 
yield can be detected by a standard $b \bar b \tau \tau$ analysis. A preliminary study 
shows that the $A^0$, $H^0 \to \tau \tau$ branching fraction can be determined to $\pm$ 15~\%
and $A^0$, $H^0 \to b \bar b$ to  $\pm$ 7~\%, from which a limit $|A_{tau}| < $ 250~GeV 
can be derived. This constrain removes the tail at low values of $\Omega_{\chi}$ and results
in a prediction of the neutralino relic density with a relative accuracy of 8~\%. A detailed 
study on full simulation to support these preliminary results is currently under way.
 
\section*{Acknowledgments}

We are grateful to Abder Djouadi for pointing out the sensitivity of the $H$ 
decay branching fractions to $A_{tau}$ and to Michael Peskin for discussion. 
This work was supported by the Director, Office of Science, of the 
U.S. Department of Energy under Contract 
No.DE-AC02-05CH11231 and used resources of the National Energy Research Scientific 
Computing Center, supported under Contract No.DE-AC03-76SF00098.


\begin{footnotesize}

\end{footnotesize}
\end{document}